\documentclass[journal,twoside,twocolumn]{IEEEtran}
\renewcommand{\IEEEQED}{}

\ifCLASSINFOpdf
\else
\fi
\usepackage{xfrac}
\usepackage{amsmath}
\usepackage{multirow}
\usepackage{color}
\usepackage{amssymb}
\usepackage{cite}
\DeclareMathOperator*{\argmin}{argmin}
\begin{document}
%
\title{Threshold-based Selective Cooperative-NOMA}
%
%
%
\author{Ferdi Kara, ~\IEEEmembership{Student Member,~IEEE,}
        Hakan Kaya 
\thanks{This work is supported by TUBITAK under 2211-E program}
\thanks{The authors are with the Department of EEE at Zonguldak Bulent Ecevit University e-mail:\{f.kara,hakan.kaya\}@beun.edu.tr.}
\vspace{-1.5\baselineskip}}
%
%
\markboth{IEEE Communications Letters,~Vol.~xx, No.~xx, Month~2019}%
{Author1 and Author2 : Threshold-based Selective Cooperative-NOMA}
%
\maketitle
\begin{abstract}

In this letter, we propose threshold-based selective cooperative-NOMA (TBS-C-NOMA) to increase the data reliability of conventional cooperative-NOMA (C-NOMA) networks. In TBS-C-NOMA, the intra-cell user forwards the symbols of cell-edge user after successive interference canceler (SIC) only if the signal-to-interference plus noise ratio (SINR) is greater than the pre-determined threshold value. Hence, the data reliability of the cell-edge user is increased by eliminating the effect of the error propagation. We derive closed-form end-to-end exact bit error probability (BEP) of proposed system for various modulation constellations. Then, the optimum threshold value is analyzed in order to minimize BEP. The obtained expressions are validated via simulations and it is revealed that TBS-C-NOMA outperforms C-NOMA and full diversity order is achieved.
\end{abstract}
\section{Introduction}
%
%
%
%
\IEEEPARstart{N}{on}-orthogonal multiple access (NOMA) is introduced to serve multiple users on the same resource block by implementing superposition coding (SC) at transmitter and iterative successive interference canceler (SIC) at receivers \cite{Saito2013}. Since NOMA provides a spectral efficient communication, the potential of NOMA for massive machine type communication (MMTC) led researchers to investigate NOMA involved systems and tremendous effort has been devoted to integrate NOMA in future wireless networks\cite{Islam2018,Ma2018}.

In NOMA systems, since the SIC is implemented at the users with better channel conditions, they have priori knowledge of the symbols of users which are former in the decoding order. Hence, the cooperation within NOMA users is proposed in \cite{Liu2015b} and the outage probability of cooperative-NOMA (C-NOMA) is analyzed. Then, the error probability of C-NOMA within two users is derived for quadrature phase shift keying (QPSK) and binary phase shift keying (BPSK) modulations \cite{Kara2019}. It is presented that error performance of C-NOMA suffers from the error propagation and the diversity order cannot be achieved. Hence, to increase data reliability of C-NOMA, in this letter, we propose threshold-based selective C-NOMA (TBS-C-NOMA) in which users with better channel conditions forwards symbols of the users with weaker channel conditions if the signal-to-interference plus noise ratio (SINR) is greater than a pre-determined threshold value. The main contributions of this paper are as follows \begin{itemize}
\item{TBS-C-NOMA network is proposed to cope with the error propagation from intra-cell user to cell-edge user in C-NOMA. Thus, end-to-end (e2e) error performance is improved and full diversity order is achieved.}
\item{The exact e2e bit error probability (BEP) of proposed TBS-C-NOMA is derived in the closed-form for different modulation constellations.}
\item{The optimum threshold value which minimizes the e2e BEP of TBS-C-NOMA is derived.}\end{itemize}
The remainder of this paper is given as follows. In section II, we introduce TBS-C-NOMA. Then, in section III, error analysis\footnote{BEP is derived for only cell-edge user since BEP of intra-cell user remains the as with conventional NOMA.} of TBS-C-NOMA is provided. The optimum threshold for TBS-C-NOMA is derived in section IV. In section V, the validation of derivations is presented via simulations. Finally, in section VI, results are discussed and the paper is concluded. 
\section{System Model}
We consider a downlink C-NOMA scheme with a base-station (BS) and two mobile users (i.e., UE1 and UE2) where each node is equipped with single antenna\footnote{Single-antenna situation is considered in this letter, however the analysis can be easily extended for multiple-antenna situations and for also different channel fadings.} as shown in Fig. 1. We assume that channel fading of each link follows $CN(0,\sigma_\lambda^2),\ \lambda=s1,s2,r$ where $s1$, $s2$ and $r$ denote the links between BS-UE1, BS-UE2 and UE1-UE2, respectively, and $\sigma_\lambda^2$ is determined by the distance (large-scale fading) between nodes. In the first phase of communication, BS implements SC and transmits the total signal of users, hence the received signals by users are given as follow:
\begin{figure}[!t]
    \centering
    \includegraphics[width=7.7cm,height=2.6cm]{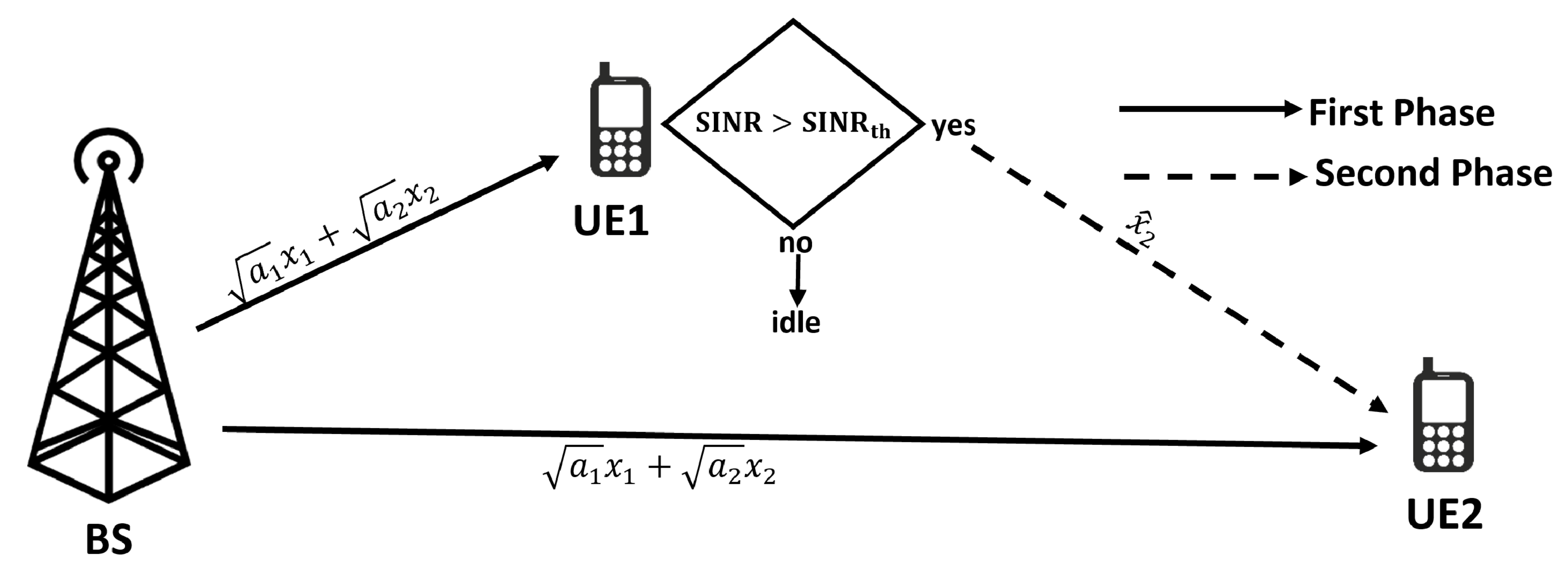}
    \caption{System model of TBS-C-NOMA}
    \label{fig1}
\end{figure}
\begin{equation}
y_\lambda=\sqrt{P_s}h_\lambda(\sqrt{a_1}x_1+\sqrt{a_2}x_2)+n_\lambda, \ \ \ \ \ \ \lambda=s1,s2
\end{equation}
where $h_\lambda$ and $n_\lambda$ denote the complex channel fading coefficient and the additive white Gaussian noise (AWGN), respectively. $n_\lambda$ is distributed as $CN(0,N_0)$. $a_1,a_2$ and $x_1,x_2$ pairs are the power allocation (PA) coefficients and the complex baseband symbols of users. Without loss of generality, UE1 and UE2 are assumed to be intra-cell user and cell-edge user according to their distances to the BS i.e., $\sigma_{s1}^2\geq\sigma_{s2}^2$, thus, $a_1<a_2$ is determined. The SINR at UE1 is given as
\begin{equation}
SINR=\frac{\rho a_2 \left|h_{s1}\right|^2}{\rho a_1\left|h_{s1}\right|^2+1}
\end{equation}
where $\rho=P_s/N_0$ is the average signal-to-noise ratio (SNR). 
In order to increase data reliability of UE2, we propose that UE1 forwards the symbols of UE2 obtained during SIC process only if the SINR is greater than a threshold value i.e., $SINR_{th}$. Hence, the received signal by UE2 in the second phase of TBS-C-NOMA is given as
\begin{equation}
y_r=\begin{cases}
0 & SINR<SINR_{th} \\
\sqrt{P_r}h_r \hat{x}_2+n_r & SINR\geq SINR_{th}
\end{cases}
\end{equation}
where $P_r$ is the power of the relay-UE1- and $h_r$ is the fading coefficient between the users. $\hat{x}_2$ denotes the detected symbols of the UE2 at UE1. The UE2 implements a maximum-ratio combining (MRC) for the received symbols in two phases and the total received symbol by UE2 is given as
\begin{equation}
y_2=y_{s1}h_{s1}^*+y_{r}h_{r}^*
\end{equation}
where $^*$ denotes complex conjugate operation. Finally, UE2 implements a maximum-likelihood (ML) detector to estimate its own symbols.
\section{Error Analysis of TBS-C-NOMA}
\subsection{The Probability SINR's being greater than the threshold}
The probability of the event that SINR is greater than the pre-determined threshold value is obtained as
\begin{equation}
\begin{split}
P_{\gamma_{s1}}^{(th)}&=P\left(SINR\geq SINR_{th}\right) \\
&=P\left(\frac{\rho a_2 \left|h_{s1}\right|^2}{\rho a_1\left|h_{s1}\right|^2+1}\geq SINR_{th}\right) \\
&=P\left(\gamma_{s1}\geq \phi_{th}\right)
\end{split}
\end{equation}
where $\gamma_{s1}=\rho\left|h_{s1}\right|^2$ and $\phi_{th}=\frac{SINR_{th}}{a_2-a_1SINR_{th}}$ are defined. We note that $a_2>a_1SINR_{th}$ condition should be accomplished otherwise UE1 always remains idle. In case the channel coefficient is Rayleigh distributed, $\gamma_{s1}$ will be exponentially distributed with the probability density function (PDF) $p_{\gamma_{s1}}(\gamma_{s1})=\sfrac{1}{\overline{\gamma}_{s1}}exp(\sfrac{-\gamma_{s1}}{\overline{\gamma}_{s1}})$, where $\overline{\gamma}_{s1}=\sigma_{s1}^2\rho$. The probability for $\gamma_{s1}\geq\phi_{th}$ is determined as
\begin{equation}
P_{\gamma_{s1}}^{(th)}=\int\limits_{\phi_{th}}^{\infty}p_{\gamma_{s1}}(\gamma_{s1})d\gamma_{s1}=exp(\sfrac{-\phi_{th}}{\overline{\gamma}_{s1}})
\end{equation}
\subsection{ Bit Error Probability of UE2 during SIC for TBS-C-NOMA}
The BEP for arbitrary modulation/fading is given as $\alpha Q(\sqrt{\beta\gamma})$. However, in case SC is implemented, the users encounter inter-user-interferences (IUI), thus the conditional BEP of the UE2 is given
\begin{equation}
P(e|_\gamma)=\sum_{i=1}^N\alpha_iQ(\sqrt{\beta_i\gamma})
\end{equation}
where $N$, $\alpha_i$ and $\beta_i$ depend on the modulation constellations of UE1 and UE2 which are given in section III.E for various constellation pairs whereas $\gamma$ is dependent on channel fading. 
Considering $SINR\geq SINR_{th}$ and by using (7), we formulate the average BEP (ABEP) of UE2 symbols at UE1 as
\begin{equation}
P_{\mathrm {II}}^{s1}(e)=\int\limits_{\phi_{th}}^{\infty}\sum_{i=1}^N\alpha_iQ(\sqrt{\beta_i\gamma_{s1}})p_{\gamma_{s1}}(\gamma_{s1})d\gamma_{s1}
\end{equation}        
By taking steps in \cite[appendix B]{Herhold}, we define
\begin{equation} 
u(\gamma)=\int p_{\gamma_{s1}}(\gamma_{s1})d\gamma_{s1}
\end{equation}
and we apply partial integration by substituting $u(\gamma)$ into (8). Then, with the aid of the Leibniz' rule \cite[eq. (0.42)]{Gradshteyn1994} and after simplifications, it yields
\begin{equation}
\begin{split}
P_{\mathrm {II}}^{s1}(e)\leq&\sum_{i=1}^N\left[\left. u(\gamma)\alpha_i Q(\sqrt{\beta_i\gamma})\right|_{\phi_{th}}^{\infty} \right. \\
&+\left.\frac{\alpha_i\sqrt{\beta_i}}{2\sqrt{2\pi}}\int\limits_{\phi_{th}}^{\infty}\frac{1}{\sqrt{\gamma}}u(\gamma) exp\left(\frac{-\beta_i}{2}\gamma\right)d\gamma\right]
\end{split}
\end{equation}
Considering channels are Rayleigh distributed and the symbol of UE2 is forwarded by UE1 only if the condition $SINR\geq SINR_{th}$ is succeeded in TBS-C-NOMA, we define
\begin{equation}
p_{\gamma_{s1}}^{(\phi_{th})}=\begin{cases}
0 & \gamma_{s1}<\phi_{th} \\
\frac{1}{exp(\sfrac{-\phi_{th}}{\overline{\gamma}_{s1}})}\frac{1}{\overline{\gamma}_{s1}}exp(\sfrac{-\gamma}{\overline{\gamma}_{s1}})& \gamma_{s1}\geq\phi_{th}
\end{cases}
\end{equation}
where $\sfrac{1}{exp(\sfrac{-\phi_{th}}{\overline{\gamma}_{s1}})}$ is the scaling factor to ensure PDF to have unit area. Substituting (11) into (9) and then into (10), the ABEP of UE2 at UE1 is derived as
\begin{equation}
\begin{split}
&P_{\mathrm {II}}^{s1}(e)\leq\sum_{i=1}^N\left[\alpha_i Q(\sqrt{\beta_i\phi_{th}})\right. \\
&-\left.\alpha_i e^{\frac{\phi_{th}}{\overline{\gamma}_{s1}}} \sqrt{\frac{1}{1+\frac{2}{\beta_i\overline{\gamma}_{s1}}}} Q\left(\sqrt{2\phi_{th}(\frac{\beta_i}{2}+\frac{1}{\overline{\gamma}_{s1}})}\right) \right]
\end{split}
\end{equation}
\subsection{Bit Error probability of Direct Transmission}
When $SINR<SINR_{th}$, the UE1 stays idle in the second phase of communication and only the direct transmission from BS to UE2 remains. In this case, with the aid of (7), the conditional BEP for symbols of UE2 is given as
\begin{equation}
P_{\mathrm {II}}^{(direct)}(e|\gamma_{s2})=\sum_{i=1}^N\alpha_iQ(\sqrt{\beta_i\gamma_{s2}})
\end{equation}
where $\gamma_{s2}=\rho\left|h_{s2}\right|^2$. The ABEP is obtained by averaging over instantaneous $\gamma_{s2}$ as $\int\limits_0^{\infty}P_{\mathrm {II}}^{(direct)}p_{\gamma_{s2}}(\gamma_{s2})d\gamma_{s2}$. The ABEP for direct transmission is derived as
\begin{equation}
P_{\mathrm {II}}^{(direct)}(e)=\sum_{i=1}^N\frac{\alpha_i}{2}\left(1-\sqrt{\frac{\beta_i\overline{\gamma}_{s2}}{2+\beta_i\overline{\gamma}_{s2}}}\right)
\end{equation}
\subsection{Bit Error probability of Diversity Transmission}
We assume that the condition $SINR\geq SINR_{th}$ is succeeded and the symbols of UE2 are detected correctly at the UE1. Since MRC is implemented at UE2 for the signals received in two phases, 
 by using (7) and utilizing \cite[eq. (14-4-11)]{Proakis1995}, the conditional BEP is given as
\begin{equation}
P_{\mathrm {II}}^{(div)}(e|_{\gamma_{s2},\gamma_r})=\sum_{i=1}^N\alpha_iQ(\sqrt{\beta_i\gamma_{s2}+2\gamma_r})
\end{equation}
where $\gamma_r=\sigma_r^2\sfrac{P_r}{N_0}$. The ABEP for diversity transmission is obtained by averaging over instantaneous $\gamma_{s2}$ and $\gamma_{r}$. It is formulated as $\int\limits_0^{\infty}\int\limits_0^{\infty}P_{\mathrm {II}}^{(div)}p_{\gamma_{s1}}(\gamma_{s1})d\gamma_{s1}p_{\gamma_{r}}(\gamma_{r})d\gamma_{r}$. By using the ABEP of 2-branch MRC \cite[eq. (14-4-15)]{Proakis1995}, the ABEP of diversity transmission turns out to be
\begin{equation}
\begin{split}
P_{\mathrm {II}}^{div}(e)=&\sum_{i=1}^N\frac{\alpha_i}{2} \left[ 1-\frac{1}{\left(\beta_i\overline{\gamma}_{s2}-2{\overline{\gamma}}_r\right)}  \right. \\
&\left.(\frac{\beta_i\overline{\gamma}_{s2}}{2}\sqrt{\frac{\beta_i\overline{\gamma}_{s2}}{2+\beta_i\overline{\gamma}_{s2}}}-{\overline{\gamma}}_r\sqrt{\frac{{\overline{\gamma}}_r}{1+{\overline{\gamma}}_r}})\right]\\
\end{split}
\end{equation}
\subsection{End-to-end (e2e) Error Performance of TBS-C-NOMA}
To obtain the e2e ABEP, we consider all the possibilities.
Hence, with the aid of the law of total probability, the e2e ABEP of TBS-C-NOMA is given as 
\begin{equation}
\begin{split}
&P_{\mathrm {II}}^{e2e}(e)=\sum_{i=1}^N (1-P_{\gamma_{s1}}^{(th)})\times P_{\mathrm {II}}^{(direct)}(e) +\\
&P_{\gamma_{s1}}^{(th)}\times\{P_{\mathrm {II}}^{div}(e)\times (1-P_{\mathrm {II}}^{s1}(e))+P_{\mathrm {II}}^{s1}(e)\times P_{\mathrm {II}}^{(prop)}(e)\}
\end{split}
\end{equation}
where $P_{prop}$ is defined as the error probability at UE2 when UE1 detects UE2's symbols erroneously and forwards to UE2. It is given for BPSK in \cite{Kara2019}, by utilizing \cite[eq.(13)]{Kara2019} and \cite[Appendix C]{Onat2008}, for M-ary constellations, we provide a tight approximation as
\begin{equation}
P_{\mathrm {II}}^{(prop)}(e)\approx\sum_{i=1}^N\alpha_i\frac{c_{j,M}\gamma_r}{\sfrac{\beta_i\gamma_{s2}}{2}+c_{j,M}\gamma_r}
\end{equation}
where
\begin{equation}
c_{j,M}\triangleq\begin{cases}
\frac{sin({\pi(2j-1)}/{M})}{sin(\sfrac{\pi}{M})} & j=1,2,\dots, \sfrac{M}{2} \\
-\frac{sin({\pi(2j+1)}/{M})}{sin(\sfrac{\pi}{M})} & j=\sfrac{M}{2}+1,\dots,M-1
\end{cases}
\end{equation}
We note that $c_{j,M}$ depends on the symbols of UE2 transmitted by BS and detected erroneously by UE1, hence we have averaged all the possibilities. The e2e ABEP of TBS-C-NOMA is derived by substituting (6), (12), (14), (16) and (18) into (17).

Throughout the paper, the analyses are provided for arbitrary modulation pairs of UE1 and UE2 which are adaptively determined according to channel qualities in wireless networks \cite{PP2016}. We provide the BEP coefficients (i.e., $N$, $\alpha_i$, $\beta_i$) for six different uncoded mode of \cite{PP2016} in Table I by utilizing the deductions of error probability for NOMA networks in \cite{Kara2018d}.
\begin{table}[t!]
\centering
   \caption{\textsc{Coefficients for error probability of UE2}}
  \label{table1}
	 \centering
  \begin{tabular}[r]{|c|c|c|}
		\hline
		\multirow{3}{*}{mode} & {Constellations} &\multirow{2}{*} {BEP} \\
		 & UE1 & \\
	&UE2&{Coefficients}\\
		\hline
	\multirow{3}{*} {1}&{BPSK}&{$N=2,\ \ \alpha_i=0.5, \quad i=1,2$} \\
		&\multirow{2}{*}{BPSK}&\multirow{2}{*}{$\beta_i= 2\left(\sqrt{a_2}\mp\sqrt{a_1}\right)^2  i=1,2$} \\
		& & \\
			\hline
			\multirow{2}{*} {2}&{BPSK}&{$N=3,\ \ \alpha_i=0.25,\ i=1,2 \quad  a_3=0.5$} \\
		&\multirow{2}{*}{QPSK}&{$\beta_i=\begin{cases} 
		2\left(\sqrt{\frac{a_2}{2}}\mp\sqrt{a_1}\right)^2 & i=1,2 \\
		a_2 &i=3\end{cases}$}\\
				\hline
			\multirow{3}{*} {3}&{QPSK}&{$N=2,\ \ \alpha_i=0.5, \quad i=1,2$} \\
		&\multirow{2}{*}{BPSK}&\multirow{2}{*}{$\beta_i= 2\left(\sqrt{a_2}\mp\sqrt{\frac{a_1}{2}}\right)^2  i=1,2$} \\
		& & \\
			\hline
			\multirow{3}{*} {4}&{QPSK}&{$N=2,\ \ \alpha_i=0.5, \quad i=1,2$} \\
		&\multirow{2}{*}{QPSK}&\multirow{2}{*}{$\beta_i= \left(\sqrt{a_2}\mp\sqrt{a_1}\right)^2  i=1,2$} \\
		& & \\
			\hline
			\multirow{2}{*} {5}&{16-QAM}&{$N=4,\ \ \alpha_i=0.25, \quad i=1,2,3,4$} \\
		&\multirow{2}{*}{BPSK}&{$\beta_i= \begin{cases}
	2\left(\sqrt{a_2}\mp\sqrt{\frac{a_1}{10}}\right)^2 & i=1,2 \\
	2\left(\sqrt{a_2}\mp3\sqrt{\frac{a_1}{10}}\right)^2& i=3,4
	\end{cases}$}  \\
						\hline
			\multirow{2}{*} {6}&{16-QAM}&{$N=4,\ \ \alpha_i=0.25, \quad i=1,2,3,4$} \\
		&\multirow{2}{*}{QPSK}&{$\beta_i= \begin{cases}
	\left(\sqrt{a_2}\mp\sqrt{\frac{a_1}{5}}\right)^2 & i=1,2 \\
	\left(\sqrt{a_2}\mp3\sqrt{\frac{a_1}{5}}\right)^2& i=3,4
	\end{cases}$}  \\
					\hline
\end{tabular}
\end{table}
\section{The Optimum Threshold for TBS-C-NOMA}
The optimum threshold is defined as the $SINR_{th}$ value which minimizes the e2e ABEP of UE2 in TBS-C-NOMA. Since the  $SINR_{th}$ has to be determined according to PA coefficients ,e.g. if $a_2<a_1SINR_{th}$, UE1 always remains idle for transmission to UE2, the optimum value which includes PA coefficient effects is given as
\begin{equation}
SINR_{th}^{opt}=\argmin_{\phi_{th}}{P_{\mathrm {II}}^{e2e}(e)}
\end{equation}
and it is derived as
\begin{equation}
\frac{dP_{\mathrm {II}}^{{e2e}}(e)}{d\phi_{th}}=0
\end{equation}
\begin{figure}[!t]
\centering
    \includegraphics[width=8cm,height=4cm]{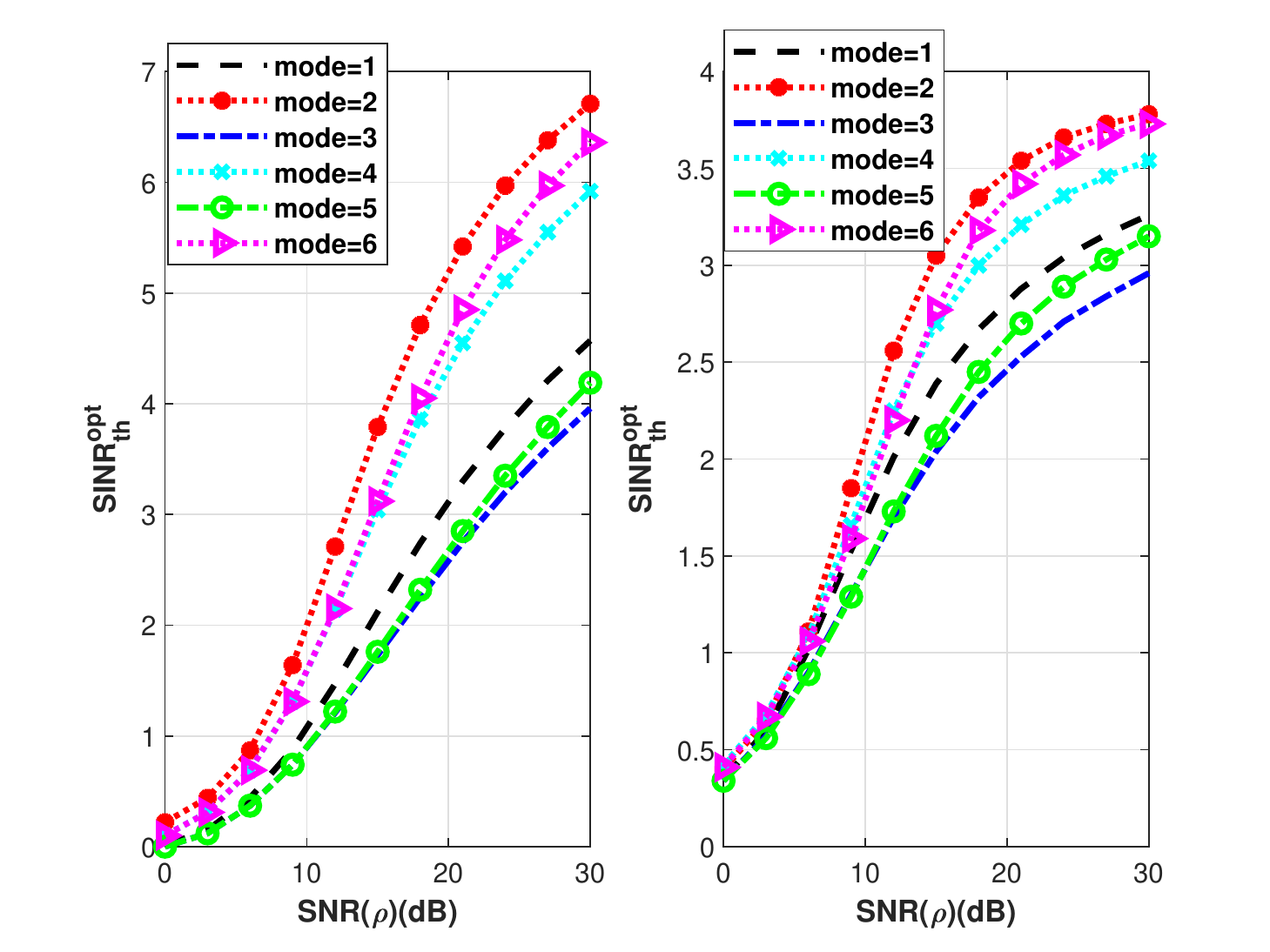}
		\centering
    \caption{Optimum threshold values a) $\sigma_{s1}^2=\sigma_{s2}^2=\sigma_{r}^2=0$dB, $a_1=0.1$, $a_2=0.9$ b) $\sigma_{s1}^2=\sigma_{r}^2=10$dB, $\sigma_{s2}^2=0$dB, $a_1=0.2$, $a_2=0.8$}
    \label{fig2}
\end{figure}
Firstly, we apply Leibniz' rule \cite[eq. (0.42)]{Gradshteyn1994} to (21).Then, with the aid of \cite{Onat2008b} and after some simplifications, the expression (21) turns out to be
\begin{equation}
\phi_{th}^{opt}=\begin{cases}
\sum\limits_{i=1}^N\frac{1}{\beta_i}\left(Q^{-1}\left(\sfrac{\delta_i}{\alpha_i}\right)\right)^2, &\delta_i <0.5 \\
0, & otherwise
\end{cases}
\end{equation}
where 
\begin{equation}
\delta_i=\frac{P_{\mathrm {II}}^{(direct)}(e|_i)-P_{\mathrm {II}}^{(div)}(e|_i)}{P_{\mathrm {II}}^{(prop)}(e|_i)-P_{\mathrm {II}}^{(div)}(e|_i)}
\end{equation}
is defined. The optimum value which minimizes the e2e ABEP is obtained by substituting (14), (16) and (18) into (22). Finally, the optimal threshold value is derived as
\begin{equation}
SINR_{th}^{opt}=\frac{a_2\phi_{th}^{opt}}{1+a_1\phi_{th}^{opt}}
\end{equation}
The $SINR_{th}^{opt}$ in (24) is given in Fig. 2 for two different scenarios when $P_r=P_s/2$. One can easily see that optimum threshold value also depends on the transmit SNR in addition to channel states and PA coefficients. To minimize the e2e ABEP of UE2, low threshold in low SNR region and high threshold in high SNR region should be implemented.
\section{Numerical Results}
In this section, the derived expressions are validated via simulations. In all figures, the lines denote the analytical curves whereas simulations are represented by markers. We assume that the power of UE1 -relay- is equal to $P_r=P_s/2$.

In Fig. 3, we present the error performance of proposed TBS-C-NOMA for all six modes in a network where $\sigma_{s1}^2=\sigma_{s2}^2=\sigma_{r}^2=0$dB. The PA coefficients are chosen as $a_1=0.1$ and $a_2=0.9$ and threshold value $SINR_{th}$ is fixed to 2. It is noteworthy that the simulations match perfectly with the analytical derivations for all modes.
\begin{figure}[!t]
    \centering
    \includegraphics[width=8cm,height=4cm]{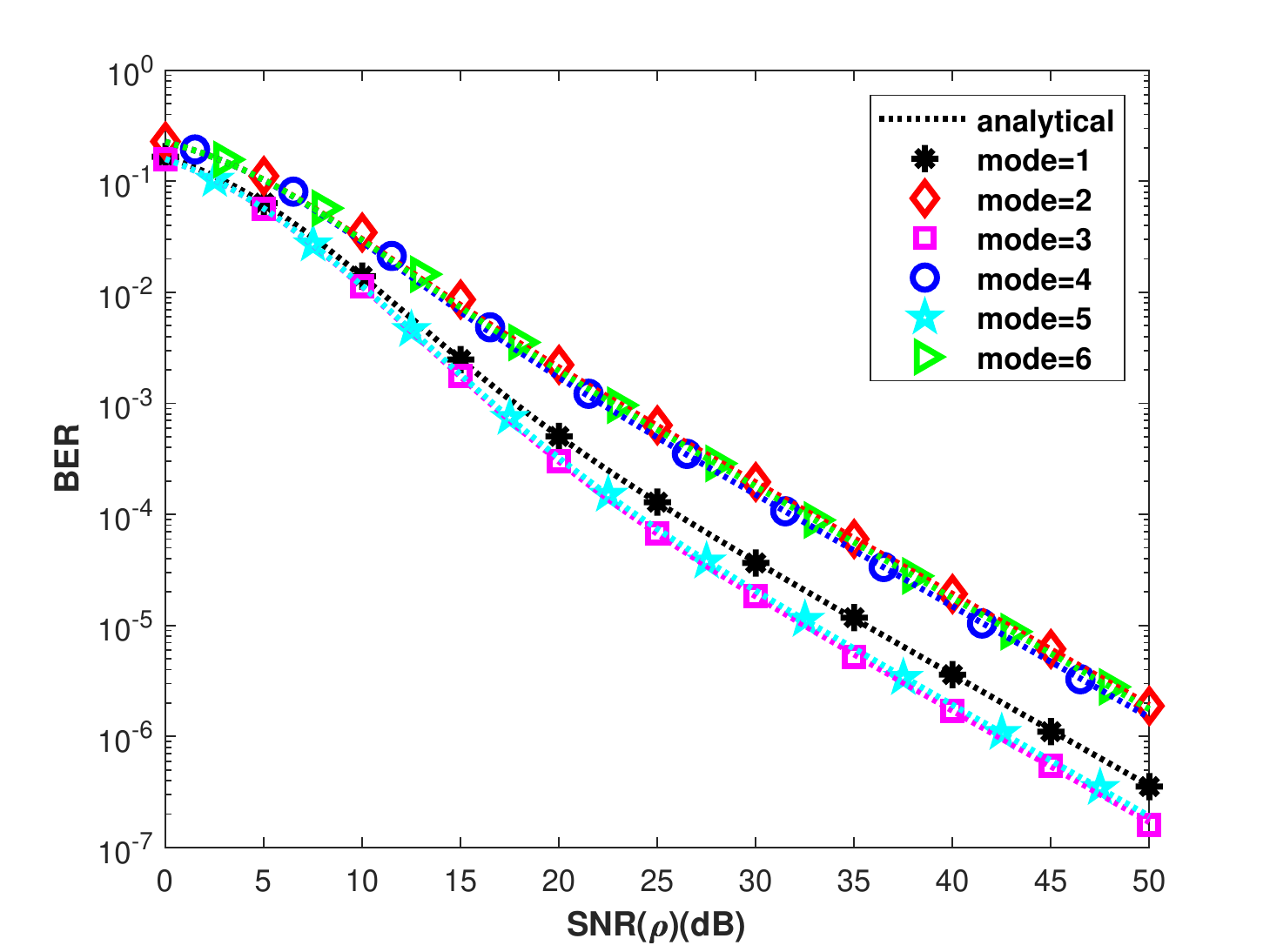}
    \caption{BER of TBS-C-NOMA for all modes  }
    \label{fig3}
\end{figure}

In Fig. 4, to emphasize the effect of the threshold, we provide the error performance of TBS-C-NOMA for different fixed-threshold values and optimum threshold value. It is assumed to be $\sigma_{s1}^2=0$dB, $\sigma_{s2}^2=\sigma_{r}^2=10$dB. We assume that mode 3 is chosen and the PA coefficients are $a_1=0.2$ and $a_2=0.8$. We also present error performances of conventional NOMA (non-cooperative), C-NOMA and C-NOMA with perfect-SIC where it is assumed that relay detects the far user's all symbols correctly (genie-aided/perfect-decoding) and no error propagation occurs. TBS-C-NOMA outperforms significantly C-NOMA since TBS-C-NOMA copes with the error propagation better than C-NOMA. Nevertheless, the error performance of TBS-C-NOMA highly depends on the threshold value. Using a low threshold value decreases the reliability of the relay -UE1- and this causes decay in the error performance in the high SNR region due to error propagation. On the other hand, using a higher threshold value causes the relay to be silent in the low SNR region, hence the error performance of TBS-C-NOMA gets worse. Nevertheless, this can be solved by the implementation of optimum threshold which is derived in section IV. In Fig. 4, one can easily see that with the use of optimum threshold TBS-C-NOMA has better error performance than the use of fixed-threshold values in whole SNR region. Hence, full diversity order is achieved and the error performance of TBS-C-NOMA gets close to C-NOMA with perfect SIC.
 \begin{figure}[!t]
    \centering
    \includegraphics[width=8cm,height=4cm]{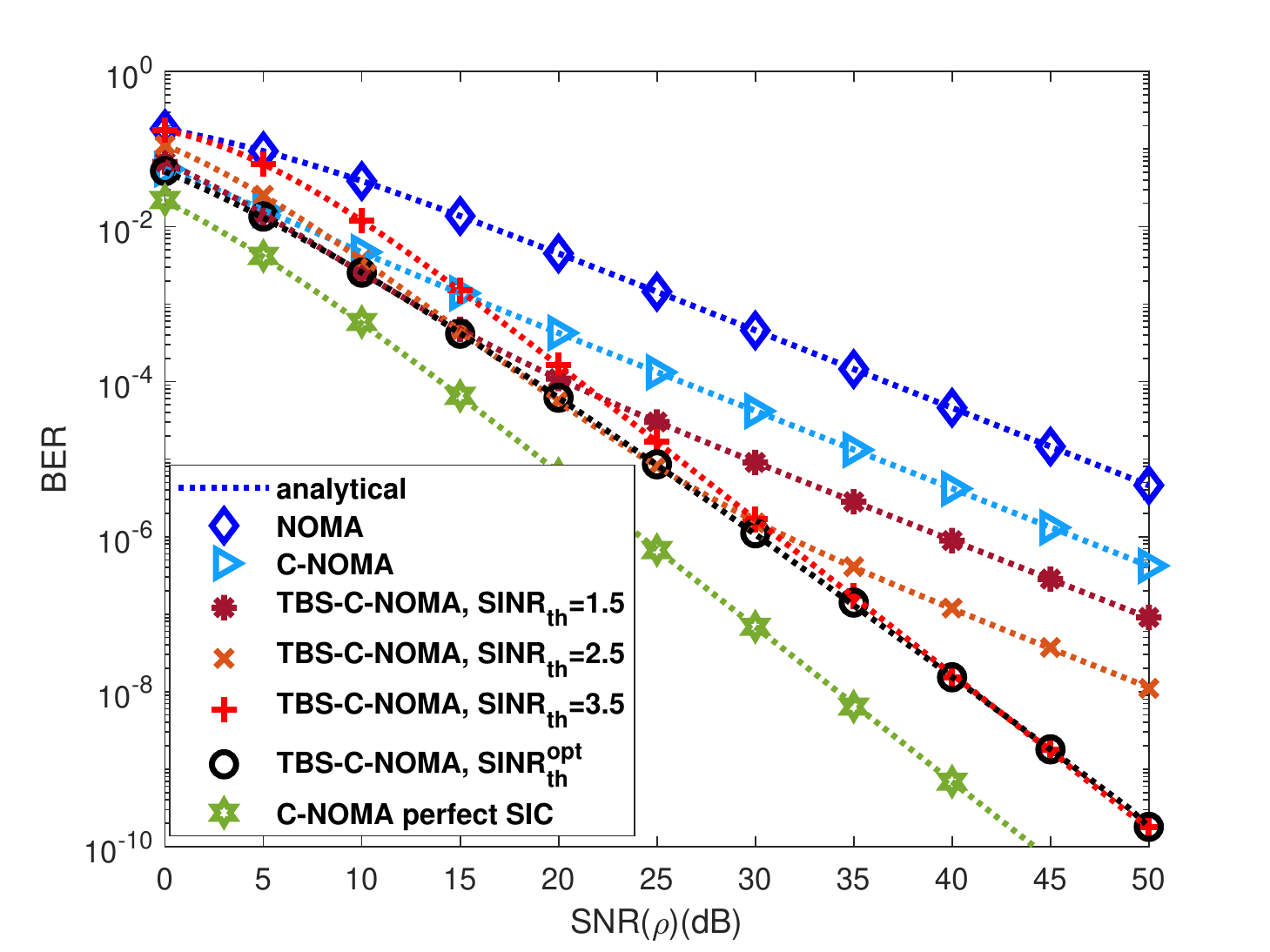}
    \caption{BER comparison of TBS-C-NOMA and C-NOMA }
    \label{fig4}
\end{figure}
\section{Conclusion}
In this letter, the TBS-C-NOMA is proposed to cope with error propagation in C-NOMA. In TBS-C-NOMA, the data reliability of cooperative phase is increased by implementing a condition that SINR of intra-cell user to be greater than a threshold value. The e2e ABEP of TBS-C-NOMA is derived for different modulation constellations. Then, the optimum threshold value is obtained in order to minimize the ABEP of TBS-C-NOMA. TBS-C-NOMA outperforms C-NOMA and full diversity order is achieved. In this letter, fixed PA is assumed. Nevertheless, optimum PA algorithms for conventional NOMA can be adapted for TBS-C-NOMA. TBS-C-NOMA provides a reliable communication for cell-edge user, and the achievable rate and outage performance of this reliable communication need to be analyzed. In addition, TBS-C-NOMA proves that introducing threshold value for such networks as NOMA and cooperative communication are involved together, can achieve better error performance. These are considered as the directions of future research.


%



\ifCLASSOPTIONcaptionsoff
  \newpage
\fi



%
\vspace{-0.75\baselineskip}
\bibliographystyle{IEEEtran}
\bibliography{tbs_c_noma} 

\begin{thebibliography}{10}
\providecommand{\url}[1]{#1}
\csname url@samestyle\endcsname
\providecommand{\newblock}{\relax}
\providecommand{\bibinfo}[2]{#2}
\providecommand{\BIBentrySTDinterwordspacing}{\spaceskip=0pt\relax}
\providecommand{\BIBentryALTinterwordstretchfactor}{4}
\providecommand{\BIBentryALTinterwordspacing}{\spaceskip=\fontdimen2\font plus
\BIBentryALTinterwordstretchfactor\fontdimen3\font minus
  \fontdimen4\font\relax}
\providecommand{\BIBforeignlanguage}[2]{{%
\expandafter\ifx\csname l@#1\endcsname\relax
\typeout{** WARNING: IEEEtran.bst: No hyphenation pattern has been}%
\typeout{** loaded for the language `#1'. Using the pattern for}%
\typeout{** the default language instead.}%
\else
\language=\csname l@#1\endcsname
\fi
#2}}
\providecommand{\BIBdecl}{\relax}
\BIBdecl

\bibitem{Saito2013}
Y.~Saito, Y.~Kishiyama, A.~Benjebbour, T.~Nakamura, A.~Li, and K.~Higuchi,
  ``{Non-Orthogonal Multiple Access (NOMA) for Cellular Future Radio Access},''
  in \emph{Proc. IEEE Veh. Technol. Conf. (VTC Spring)}, jun 2013, pp. 1--5.

\bibitem{Islam2018}
S.~M. Islam, M.~Zeng, O.~A. Dobre, and K.~S. Kwak, ``{Resource Allocation for
  Downlink NOMA Systems: Key Techniques and Open Issues},'' \emph{IEEE Wirel.
  Commun.}, vol.~25, no.~2, pp. 40--47, 2018.

\bibitem{Ma2018}
G.~Ma, B.~Ai, F.~Wang, X.~Chen, Z.~Zhong, Z.~Zhao, and H.~Guan, ``{Coded Tandem
  Spreading Multiple Access for Massive Machine-Type Communications},''
  \emph{IEEE Wirel. Commun.}, vol.~25, no.~2, pp. 75--81, 2018.

\bibitem{Liu2015b}
Z.~Ding, M.~Peng, and H.~V. Poor, ``{Cooperative Non-Orthogonal Multiple Access
  in 5G Systems},'' \emph{IEEE Commun. Lett.}, vol.~19, no.~8, pp. 1462--1465,
  aug 2015.

\bibitem{Kara2019}
F.~Kara and H.~Kaya, ``{On the Error Performance of Cooperative-NOMA with
  Statistical CSIT},'' \emph{IEEE Commun. Lett.}, vol.~23, no.~1, pp. 128--131,
  2019.

\bibitem{Herhold}
P.~Herhold, E.~Zimmermann, and G.~Fettweis, ``{A simple cooperative extension
  to wireless relaying},'' in \emph{Proc. Int. Zurich Semin. Commun.}, feb
  2004, pp. 36--39.

\bibitem{Gradshteyn1994}
I.~Gradshteyn and I.~Ryzhik, \emph{{Table of Integrals, Series, and Products}},
  5th~ed.\hskip 1em plus 0.5em minus 0.4em\relax San Diego: CA: Academic Press,
  feb 1994.

\bibitem{Proakis1995}
J.~G. Proakis, \emph{{Digital Communications}}, 3rd~ed.\hskip 1em plus 0.5em
  minus 0.4em\relax New York: McGraw-Hill, 1995.

\bibitem{Onat2008}
F.~A. Onat, A.~Adinoyi, Y.~Fan, H.~Yanikomeroglu, J.~S. Thompson, and I.~D.
  Marsland, ``{Threshold selection for SNR-based selective digital relaying in
  cooperative wireless networks},'' \emph{IEEE Trans. Wirel. Commun.}, vol.~7,
  no.~11, pp. 4226--4237, 2008.

\bibitem{PP2016}
3GPP, ``{RP-160680:Downlink Multiuser Superposition Transmission for LTE},''
  Tech. Rep., 2016.

\bibitem{Kara2018d}
F.~Kara and H.~Kaya, ``{BER performances of downlink and uplink NOMA in the
  presence of SIC errors over fading channels},'' \emph{IET Commun.}, vol.~12,
  no.~15, pp. 1834--1844, sep 2018.

\bibitem{Onat2008b}
F.~A. Onat, A.~Adinoyi, Y.~Fan, H.~Yanikomeroglu, and J.~S. Thompson,
  ``{Optimum Threshold for SNR-Based Selective Digital Relaying Schemes in
  Cooperative Wireless Networks},'' in \emph{Proc. IEEE Wirel. Commun. Netw.
  Conf.}, mar 2007, pp. 969--974.

\end{thebibliography}

%






\end{document}